\documentclass[prl,print,twocolumn,superscriptaddress,showpacs]{revtex4}
\usepackage[dvips]{graphics,color}
\usepackage{amsfonts}
\usepackage{amsmath}
\usepackage{amssymb}
\usepackage{graphicx}

\input{tcilatex}
\begin{document}

\title{Spin-Parity Effect in Violation of Bell's Inequalities}
\author{Zhigang Song}
\affiliation{Institute of Theoretical Physics and Department of Physics, Shanxi
University, Taiyuan, Shanxi 030006, China}
\author{J. -Q. Liang}
\thanks{jqliang@sxu.edu.cn}
\affiliation{Institute of Theoretical Physics and Department of Physics, Shanxi
University, Taiyuan, Shanxi 030006, China}
\author{L. -F. Wei}
\affiliation{State Key Laboratory of Optoelectronic Materials and Technologies, School of
Physics and Engineering, Sun Yat-Sen University, Guangzhou 510275, China}
\affiliation{Quantum Optoelectronics Laboratory, School of Physics and Technology,
Southwest Jiaotong University, Chengdu 610031, China}

\begin{abstract}
Analytic formulas of Bell correlations are derived in terms of quantum
probability statistics under the assumption of measuring
outcome-independence and the Bell inequalities (BIs) are extended to general
bipartite-entanglement states of arbitrary spins. For a spin-$1/2$ entangled
state we find analytically that the violations of BIs really resulted from
the quantum non-local correlations. However, the BIs are always satisfied
for the spin-$1$ pure entangled-states. More generally the quantum
non-locality does not lead to the violation for the integer spins since the
non-local interference effects cancel each other by the quantum
statistical-average. Such a cancellation no longer exists for the
half-integer spins due to the nontrivial Berry phase, and thus the violation
of BIs is understood remarkably as an effect of geometric phase.
Specifically, our generic observation of the spin-parity effect can be
experimentally tested with the entangled photon-pairs.
\end{abstract}

\maketitle


\section{Introduction}

Non-locality is one of the most peculiar characteristic of quantum mechanics
beyond our intuition of space and time, which has no classical
correspondence and thus can coexist with relativistic causality. Bohm
established a formulation known as the hidden-variable model, which shows
the non-local nature of quantum mechanics explicitly by the quantum
potential, and thus the quantum mechanics is not able to be reproduced by a
local theory. Quantum entangled-state of two particles separated by a
space-like interval is the typical example of non-locality, which although
was originally considered by Einstein, Podolsky, and Rosen (EPR) as an
evidence of counterexample for the incompleteness of quantum mechanics, has
become a key ingredient of applications in quantum computation and
communication \cite{3,4}.

The pioneering work of Bell stimulated by the EPR argument in the spin
version provides a possibility of quantitative test for non-local
correlations. The well known Bell's inequality (BI), which serves as a
criteria of classical action, is derived in terms of classical statistics
models with assumptions of locality and outcome-independence. Based on the
same logic, a more suitable version of the inequality for experimental test
is formulated by Clauser-Horne-Shimony-Holt (CHSH). Until now, BI is shown
to be violated with various quantum non-localities, and the overwhelming
experimental evidence for the violation \cite{9,12,13,14,15,16,17,18,19,20}
opens up the most intriguing aspect of non-locality in quantum mechanics. BI
was also reformulated in a straightforward way by Wigner with spin
alignment-measurements along only one direction. Really, non-locality has
become a powerful resource \cite{4,23} of quantum information science \cite%
{24,25,26,28,29,31,32,34,35}, wherein the underlying physical-principle is
obscure \cite{22}.\textbf{\ }A new type of inequality known as Leggett
inequality \cite{35.1} was considered as the generation of BI to non-local
deterministic models according to Ref.\cite{9}, which has become an active
research topic both theoretically and experimentally in recent years \cite%
{3,4},\cite{17,18,19,20,23}.

We in this paper adopt the density operators of entangled states to evaluate
the outcome-correlations in terms of quantum probability-statistics. Since
the density operators can be separated to local and non-local parts, the
non-local outcome-correlations are obtained explicitly. Note that BI, which
is a constraint on experimental statistics for any locally causal theory,
usually admits a local model (LM) \cite{36} and is based on classical
statistics. Here we present analytic quantum-correlations for both the pure
entangled-states (PES) and complete mixed-states. The former is non-local,
while the latter is a LM. It is demonstrated that the inequalities indeed
should be obeyed by the LM and, remarkably their violations in the PESs are
seen to be the effects of Berry phase (BP).\ It is commonly believed that
there exist two kinds of non-localities: one is considered in the sense of
BI violation and the other is called the dynamic non-locality in relation
with the topological phases (i.e. the well-known Aharonov-Bohm and Berry
phases) of quantum states. They seem completely different in physical
principles. We in this paper show for the first time that these two
non-local effects are closely related and moreover the violation of BI
directly resulted from the geometric phase, based on which a spin-parity
effect is predicted.

The spin-parity phenomenon will be firstly demonstrated in terms of the
spin-singlet states following the original model of Bell, however, with the
generation to arbitrary high-spins. Then we extend the BI and its violation
to all spin-states of bipartite-entanglement with two-dimensional
measurement-outcomes. The high-dimensional correlation measurements for
entangled angular-momentum states have been extensively investigated in
Refs. \cite{37,38,41}. The violation of BIs was theoretically verified long
ago in a clever manner for spin-$1/2$ .\cite{42} and spin-$s$ entangled
states \cite{43}. We compare in section IV the spin-parity phenomenon with
the general result of violation in Refs.\cite{42,43} showing the consistence
between them.

\section{Violation of Bell's inequalities for spin-1/2 while nonviolation
for spin-1 pure entangled-states}

\subsection{\protect\bigskip Spin-$1/2$ singlet state}

We begin with a spin-$1/2$ singlet state whose density-operator $\overset{%
\wedge }{\rho }$ can be split into two parts%
\begin{equation*}
\overset{\wedge }{\rho }=\overset{\wedge }{\rho }_{lc}+\overset{\wedge }{%
\rho }_{nlc}
\end{equation*}%
in order to see the non-local quantum effect explicitly\textbf{.} Here, the
complete mixed state $\overset{\wedge }{\rho }_{lc}=\left( |+,-\rangle
\langle +,-|+|-,+\rangle \langle -,+|\right) /2$ (with $\widehat{\sigma }%
_{z}|\pm \rangle =\pm |\pm \rangle $) is called the LM or Bell model,
wherein the two particles are independent (without entanglement) and their
behaviors are ruled by local theory. While, the second term $\overset{\wedge 
}{\rho }_{nlc}=-\left( |+,-\rangle \langle -,+|+|-,+\rangle \langle
+,-|\right) /2$ denotes the non-local interference, which remains even if
the two particles are separated in a space-like interval. Following Bell the
measurements of two spins are performed independently along arbitrary
directions ${\mathbf{a}}$ and ${\mathbf{b}}$, respectively. The measuring
outcomes are the eigenvalues of projection spin-operators $\widehat{\sigma }%
\cdot {\mathbf{a}}$ and $\widehat{\sigma }\cdot {\mathbf{b}}$, i.e., %
$\hat{\sigma}\cdot \mathbf{a|\pm a}\rangle =\pm \mathbf{|\pm a}\rangle $ and 
$\hat{\sigma}\cdot \mathbf{b|\pm b}\rangle =\pm \mathbf{|\pm b}\rangle $, 
according to the quantum measurement theory. Above, 
\begin{equation*}
\mathbf{|+r}\rangle \mathbf{=}\cos (\frac{\theta _{r}}{2})|+\rangle +\sin (%
\frac{\theta _{r}}{2})e^{i\phi _{r}}|-\rangle
\end{equation*}%
and 
\begin{equation*}
\mathbf{|-r}\rangle \mathbf{=}\sin (\frac{\theta _{r}}{2})|+\rangle -\cos (%
\frac{\theta _{r}}{2})e^{i\phi _{r}}|-\rangle
\end{equation*}%
are the spin coherent states. Here, $\mathbf{r=(}\sin \theta _{r}\cos \phi
_{r},\sin \theta _{r}\sin \phi _{r}$,$\cos \theta _{r})$ with $\mathbf{r}=%
\mathbf{a},\mathbf{b}$ is an unit vector parametrized by the polar and
azimuthal angles ($\theta _{r}$ and $\phi _{r}$) in the coordinate frame
with $z$-axis along the direction of the initial spin-polarization. The
outcome-independent base vectors of two-particle measurements are the
product eigenstates of operators $\hat{\sigma}\cdot \mathbf{a}$ and $\hat{%
\sigma}\cdot \mathbf{b}$. They are denoted as: 
\begin{equation}
|1\rangle \!\!=\!\!\mathbf{|\!\!+\!a,\!\!+\!b}\rangle ,\!|2\rangle \!\!=\!\!%
\mathbf{|\!\!+\!a,\!\!-\!b}\rangle ,\!|3\rangle \!\!=\!\!\mathbf{%
|\!\!-\!a,\!+\!b}\rangle ,|4\rangle \!\!=\!\!\mathbf{|\!\!-\!a,\!\!-\!b}%
\rangle ,
\end{equation}%
in which the $4\times 4$ density-matrix elements $\rho _{ij}=\langle i|%
\overset{\wedge }{\rho }|j\rangle $, ($i,j=1,2,3,4$) of the two-spin singlet
state $\hat{\rho}$ are evaluated, and the nonvanishing matrix elements of
correlation operator 
\begin{equation*}
\hat{\Omega}(ab)=(\hat{\sigma}\cdot \mathbf{a)}(\hat{\sigma}\cdot \mathbf{b)}
\end{equation*}%
are obviously 
\begin{equation*}
\Omega _{11}(ab)=\Omega _{44}(ab)=1
\end{equation*}%
and 
\begin{equation*}
\Omega _{22}(ab)=\Omega _{33}(ab)=-1.
\end{equation*}%
The correlation probability is thus obtained in terms of the quantum
statistical-average:%
\begin{equation}
P(ab)=Tr[\hat{\Omega}(ab)(\hat{\rho}_{lc}+\hat{\rho}_{nlc})]=\rho _{11}+\rho
_{44}-\rho _{22}-\rho _{33}
\end{equation}%
Here, the explicit forms of density-matrix elements are given by 
\begin{equation*}
\rho _{11}=\rho _{11}^{lc}+\rho _{11}^{nlc}=\rho _{44},
\end{equation*}%
and$\quad $%
\begin{equation*}
\rho _{22}=\rho _{22}^{lc}+\rho _{22}^{nlc}=\rho _{33},
\end{equation*}%
with 
\begin{equation*}
\rho _{11}^{lc}=\rho _{44}^{lc}=\frac{1}{2}(K_{a}^{2}\Gamma
_{b}^{2}+K_{b}^{2}\Gamma _{a}^{2}),
\end{equation*}%
\begin{eqnarray}
\rho _{11}^{nlc} &=&\rho _{44}^{nlc}=-\rho _{22}^{nlc}=-\rho _{33}^{nlc} \\
&=&-\frac{1}{4}\sin \theta _{a}\sin \theta _{b}\cos (\phi _{a}-\phi _{b}), 
\notag
\end{eqnarray}%
and 
\begin{equation*}
\rho _{22}^{lc}=\rho _{33}^{lc}=\frac{1}{2}(K_{a}^{2}K_{b}^{2}+\Gamma
_{a}^{2}\Gamma _{b}^{2}).
\end{equation*}%
Above, $K_{r}^{m}=\cos ^{m}(\theta _{r}/2)$ and $\Gamma _{r}^{m}=\sin
^{m}(\theta _{r}/2)$, with $r=a,b,c,d;$ $m=1,2,3,\cdot \cdot \cdot $.
Consequently, the outcome correlation given by Eq. (2) can also be separated
into the local and non-local parts, i.e., 
\begin{equation*}
P(ab)=P_{lc}(ab)+P_{nlc}(ab),
\end{equation*}%
with the local one 
\begin{equation}
P_{lc}(ab)=Tr[\hat{\Omega}(ab)\hat{\rho}_{lc}]=-\cos \theta _{a}\cos \theta
_{b},
\end{equation}%
and the non-local one: 
\begin{equation*}
P_{nlc}(ab)=-\sin \theta _{a}\sin \theta _{b}\cos (\phi _{a}-\phi _{b}).
\end{equation*}

In terms of the local correlation-probabilities Eq. (4), we now demonstrate
how the original BI, 
\begin{equation*}
|P_{lc}(ab)-P_{lc}(ac)|\leq 1+P_{lc}(bc),
\end{equation*}%
and the CHSH inequality 
\begin{equation*}
P_{CHSH}^{lc}\leq 2,
\end{equation*}%
can be verified perfectly, where 
\begin{equation*}
P_{\mathrm{CHSH}}^{lc}=|P_{lc}(ab)+P_{lc}(ac)+P_{lc}(db)-P_{lc}(dc)|
\end{equation*}%
is the combined correlation-probability of four-direction measurements.
After simple algebras we obtain the two sides of BI being 
\begin{eqnarray*}
1+P_{lc}(bc) &=&\sin ^{2}(\frac{\theta _{b}+\theta _{c}}{2})+\sin ^{2}(\frac{%
\theta _{b}-\theta _{c}}{2}), \\
P_{lc}(ab)-P_{lc}(ac) &=&-2\cos \theta _{a}\sin (\frac{\theta _{b}+\theta
_{c}}{2})\sin (\frac{\theta _{b}-\theta _{c}}{2}),
\end{eqnarray*}%
respectively, from which the BI is obvious satisfied for arbitrary
directions $\mathbf{a}$, $\mathbf{b}$ and $\mathbf{c}$. Similarly,
substitution of the local correlation formula Eq. (4) into $P_{\mathrm{CHSH}%
}^{lc}$ yields directly the CHSH inequality 
\begin{equation*}
P_{\mathrm{CHSH}}^{lc}=|\cos \theta _{a}(\cos \theta _{b}+\cos \theta
_{c})+\cos \theta _{d}(\cos \theta _{b}-\cos \theta _{c})|\leq 2.
\end{equation*}%
Secondly, for the PES $\overset{\wedge }{\rho }$ we obtain the well-known
quantum correlation-probability 
\begin{equation*}
P(ab)=-\mathbf{a}\cdot \mathbf{b}.
\end{equation*}%
The violation of BI with $P(ab)$ has been investigated extensively. Indeed,
for example, we let\textbf{\ }vector\textbf{\ }$\mathbf{b}$ be perpendicular
to\textbf{\ }$\mathbf{c}$, then $P(bc)=-\mathbf{b\cdot c}=0$\textbf{\ }and
the greater side of BI is\textbf{\ }$1+P(bc)=1$\textbf{. }On the other hand,
if the vector\textbf{\ }$\mathbf{a}$\textbf{\ }is parallel to\textbf{\ }$%
\mathbf{b}-\mathbf{c},$ the less side\textbf{\ }$|P(ab)-P(ac)|=|\mathbf{a}%
\cdot (\mathbf{b}-\mathbf{c})|=\sqrt{2}$ becomes greater than the\textbf{\ }$%
1+P(bc)$\textbf{. }Furthermore, the CHSH inequality becomes the known
maximum inequality form 
\begin{equation*}
P_{CHSH}=\!\!|\mathbf{a}\cdot (\mathbf{b}+\mathbf{c})+\mathbf{d}\cdot (%
\mathbf{b}-\mathbf{c})|\!\!\leq \!\!2\sqrt{2}.
\end{equation*}%
\textbf{\ }

The local correlation Eq. (4) can also be applied to the so-called\textbf{\ }%
Wigner inequality, in which only one direction of spin-polarization is
measured. In this case,\textbf{\ }correlation-probability becomes\textbf{\ }%
\begin{equation*}
P_{lc}(+a,+b)=\rho _{11}^{lc}=\rho _{44}^{lc}\mathbf{.}
\end{equation*}%
Using the formula of density-matrix element $\rho _{11}^{lc}$ we obtain%
\textbf{\ }%
\begin{eqnarray*}
&&P_{lc}(+a,+b)+P_{lc}(+a,+c) \\
&=&1-\cos \theta _{a}\cos (\frac{\theta _{b}+\theta _{c}}{2})\cos (\frac{%
\theta _{b}-\theta _{c}}{2}),
\end{eqnarray*}%
and 
\begin{equation*}
P_{lc}(+b,+c)=1-[\cos ^{2}(\frac{\theta _{b}+\theta _{c}}{2})+\cos ^{2}(%
\frac{\theta _{b}-\theta _{c}}{2})].
\end{equation*}%
Thus, the Wigner inequality\textbf{\ }%
\begin{equation*}
P_{lc}(+b,+c)\leq P_{lc}(+a,+b)+P_{lc}(+a,+c)
\end{equation*}%
is satisfied obviously for the LM.

\subsection{Spin-$1$ singlet}

For the spin-1 singlet the spin-coherent states are found as%
\begin{equation*}
|+\mathbf{a}\rangle _{1}=K_{a}^{2}|+1\rangle +\frac{1}{\sqrt{2}}\sin \theta
_{a}e^{i\phi _{a}}|0\rangle +\Gamma _{a}^{2}e^{i2\phi _{a}}|-1\rangle
\end{equation*}%
and 
\begin{equation*}
|-\mathbf{a}\rangle _{1}=\Gamma _{a}^{2}|+1\rangle -\frac{1}{\sqrt{2}}\sin
\theta _{a}e^{i\phi _{a}}|0\rangle +K_{a}^{2}e^{i2\phi _{a}}|-1\rangle ,
\end{equation*}%
where $\hat{s}_{z}|\pm 1\rangle =\pm |\pm 1\rangle $. The
correlation-measurement probability of two spin-$1$ particles can be
obtained by using the following matrix elements: 
\begin{eqnarray*}
\rho _{(1)11}^{lc} &=&\rho _{(1)44}^{lc}=\frac{1}{2}(K_{a}^{4}\Gamma
_{b}^{4}+K_{b}^{4}\Gamma _{a}^{4}), \\
\rho _{(1)11}^{nlc} &=&\rho _{(1)44}^{nlc}=-\frac{1}{32}[\sin ^{2}\theta
_{a}\sin ^{2}\theta _{b}\cos 2(\phi _{a}-\phi _{b})], \\
\rho _{(1)22}^{lc} &=&\rho _{(1)33}^{lc}=\frac{1}{2}(K_{a}^{4}K_{b}^{4}+%
\Gamma _{a}^{4}\Gamma _{b}^{4}),
\end{eqnarray*}%
and%
\begin{equation}
\rho _{(1)22}^{nlc}=\rho _{(1)33}^{nlc}=\rho _{(1)11}^{nlc}=\rho
_{(1)44}^{nlc}.  \label{5}
\end{equation}%
Obviously, the equality feature of the non-local parts Eq. (5) is different
from that of Eq. (3) for the spin-$1/2$ case, where the reversal of
spin-polarization measurements leads to a minus sign, which is due to the
nontrivial BP. In the spin-$1$ case the BP factor is only a trivial one $%
e^{i2\pi }=1$, which will be explained in the following section. Therefore,
the contributions of non-local interference between the same and opposite
spin-polarizations cancel each other and the outcome correlation is only
originated from the local part $\hat{\rho}_{(1)}^{lc}$ , i.e., 
\begin{equation*}
P_{(1)}(ab)=P_{(1)}^{lc}(ab).
\end{equation*}%
\emph{Thus, both the Bell and CHSH inequalities will not be violated for the
spin-}$1$\emph{\ case. }We now extend this novel observation, that the BIs
are violated by the spin-$1/2$ but not the spin-$1$ PESs, to the arbitrarily
high spins.

\section{Berry phase and spin-parity phenomenon for spin-$S$ singlet states}

For the high spin-$s$ particles, the two-spin singlet is the so-called Bell
cat-state whose density operator can also be separated into two parts: the
complete mixed $\hat{\rho}_{(s)}^{lc}=(|+s,-s\rangle \langle
+s,-s|+|-s,+s\rangle \langle -s,+s|)/2$ and the non-local ones $\hat{\rho}%
_{(s)}^{_{nlc}}=-(|+s,-s\rangle \langle -s,+s|+|-s,+s\rangle \langle
+s,-s|)/2$, respectively. Here, $|\pm s\rangle $ are called the macroscopic
quantum states wherein the minimum uncertainty relation $|\langle \widehat{s}%
_{z}\rangle |=2\langle (\Delta \widehat{s}_{x})^{2}\rangle ^{1/2}\langle
(\Delta \widehat{s}_{y})^{2}\rangle ^{1/2}$ is satisfied. So that the Bell
cat is actually the entangled Schr\H{o}dinger cat-state. We assume that the
measurements are restricted on the macroscopic states, namely the spin
coherent states $\mathbf{|\pm a}\rangle _{s}:\hat{s}\cdot \mathbf{a|\pm a}%
\rangle _{s}=\pm s\mathbf{|\pm a}\rangle _{s}$. These spin coherent states
can be generated from the extreme states $|\pm s\rangle $ such that%
\begin{equation*}
\mathbf{|\pm a}\rangle _{s}=\hat{R}|\pm s\rangle
\end{equation*}%
with the generation operator 
\begin{equation*}
\hat{R}=e^{i\theta _{a}\mathbf{m}\cdot \widehat{s}}.
\end{equation*}%
The unit-vector $\mathbf{m}$ in the $x-y$ plane is perpendicular to the
plane expanded by $z$-axis and the unit vector $\mathbf{a}$.

The explicit forms of spin coherent-states in the representation of Dicke
states are given by \cite{39} 
\begin{equation*}
|+\mathbf{a}\rangle _{s}=\sum_{m=-s}^{s}\binom{2s}{s+m}\!\!^{\frac{1}{2}%
}K_{a}^{s+m}\Gamma _{a}^{s-m}e^{i(s-m)\phi _{a}}|m\rangle ,
\end{equation*}%
\begin{equation*}
|-\mathbf{a}\rangle _{s}=\sum_{m=-s}^{s}\binom{2s}{s+m}^{\frac{1}{2}%
}K_{a}^{s-m}\Gamma _{a}^{s+m}e^{i(s-m)(\phi _{a}+\pi )}|m\rangle .
\end{equation*}%
In the outcome-independent base vectors of two-particle measurements
corresponding to Eq. (1), however, with the spin-$1/2$ replaced by $s$, the
density-matrix elements for the Bell cat-state are given by: 
\begin{eqnarray*}
\rho _{(s)11}^{lc} &=&\rho _{(s)44}^{lc}=\frac{1}{2}(K_{a}^{4s}\Gamma
_{b}^{4s}+K_{b}^{4s}\Gamma _{a}^{4s}), \\
&& \\
\rho _{(s)11}^{nlc} &=&\rho _{(s)44}^{nlc}=-K_{a}^{2s}\Gamma
_{a}^{2s}K_{b}^{2s}\Gamma _{b}^{2s}\cos [2s(\phi _{a}-\phi _{b})], \\
&& \\
\rho _{(s)22}^{lc} &=&\rho _{(s)33}^{lc}=\frac{1}{2}(K_{a}^{4s}K_{b}^{4s}+%
\Gamma _{a}^{4s}\Gamma _{b}^{4s}),
\end{eqnarray*}%
and

\begin{equation}
\rho _{(s)22}^{nlc}=\rho _{(s)33}^{nlc}=(-1)^{2s}\rho _{(s)11}^{nlc}.
\end{equation}%
It is noted that, each of the non-local elements $\rho _{(s)22}^{nlc}$ and $%
\rho _{(s)33}^{nlc}$ for the opposite spin-polarizations, e.g., $\rho
_{_{(s)}22}^{nlc}=-[\langle -s|+\mathbf{a}\rangle \langle +\mathbf{a}%
|+s\rangle \langle +s|-\mathbf{b}\rangle \langle -\mathbf{b}|-s\rangle
\break +\langle +s|+\mathbf{a}\rangle \langle +\mathbf{a}|-s\rangle \langle
-s|-\mathbf{b}\rangle \langle -\mathbf{b}|+s\rangle ]/2$, possesses an
additional BP factor $(-1)^{2s}$ compared with the same polarization (i.e. $%
\rho _{(s)11}^{nlc}=\rho _{(s)44}^{nlc}$). It resulted from the reversal of
spin-polarization measurements and can be directly evaluated from the above
inner products of spin coherent states in the non-local elements. In fact,
the BP factor $e^{isA}$ naturally exists in the inner product of two
spin-coherent-states \cite{39,40}, e.g., 
\begin{equation*}
\langle \mathbf{+n}_{1}|+\mathbf{n}_{2}\rangle =\left( \frac{1-\mathbf{n}%
_{1}\cdot \mathbf{n}_{2}}{2}\right) ^{\frac{1}{2}}e^{isA},
\end{equation*}%
where $A$ is an area expanded by the unite vectors $\mathbf{n}_{1}$, $%
\mathbf{n}_{2}$ and the north pole on the unite sphere.

Obviously, the non-local elements $\rho _{(s)22}^{nlc}$ and $\rho
_{(s)33}^{nlc}$ (for the opposite spin-polarization measurements) differ
from $\rho _{(s)11}^{nlc}$= $\rho _{(s)44}^{nlc}$ (of the same direction
measurements) only by the BP factor $(-1)^{2s}$, seen from Eq. (6). As a
consequence, the contributions of non-local elements in the outcome
correlation $P_{(s)}(a,b)$ [i.e. the spin-$s$ version of Eq. (2)] cancel
each other in the integer-spin Bell cat-state, since the BP factor is
trivial and thus\textbf{\ }$\rho _{(s)11}^{nlc}=\rho _{(s)22}^{nlc}$.
However, in the half-integer\textbf{\ }spin\textbf{\ }case an additional
minus sign of the BP factor leads to\textbf{\ }$\rho _{(s)11}^{nlc}=-\rho
_{(s)22}^{nlc}$\textbf{\ }and thus the non-local part of correlation
probability\textbf{\ }$P_{(s)}^{nlc}(ab)$\textbf{\ }does not vanish\textbf{.}
Based on these analysis, we may reach a general spin parity phenomenon for
the PES: \textit{\ The measured two-particle correlations for the Bell
cat-states of integer spins should satisfy the inequality predicted by the
usual local theory, and the violation of BI for the half-integer spins is a
direct result of the BP.}

On the other hand, with the above density-matrix elements the normalized
(the correlation per spin value) Bell correlations for the\textbf{\ }LM read%
\begin{equation}
P_{(s)}^{lc}(ab)=-(K_{a}^{4s}-\Gamma _{a}^{4s})(K_{b}^{4s}-\Gamma _{b}^{4s}).
\label{7}
\end{equation}%
In terms of which it is straightforward to prove that the CHSH inequality 
\begin{equation*}
\left\vert
P_{(s)}^{lc}(ab)+P_{(s)}^{lc}(ac)+P_{(s)}^{lc}(db)-P_{(s)}^{lc}(dc)\right%
\vert \leq 2
\end{equation*}%
is always satisfied. Moreover, a new inequality%
\begin{equation}
4|P_{(s)}^{lc}(bc)|\!\!\geq
\!\![P_{(s)}^{lc}(ab)\!\!+\!\!P_{(s)}^{lc}(ac)]^{2}\!\!-\!%
\![P_{(s)}^{lc}(ab)\!\!-\!\!P_{(s)}^{lc}(ac)]^{2}
\end{equation}%
is found to be an universal result for arbitrary spins. Certainly, all these
inequalities should be satisfied by not only the LM $\hat{\rho}_{lc}$ of
integer and half-integer spins but also the Bell cat-states $\hat{\rho}%
_{(s)} $ of integer spins. As a matter of fact, with the correlation Eq. (7)
the BI becomes 
\begin{eqnarray*}
&&1-(K_{b}^{4s}-\Gamma _{b}^{4s})(K_{c}^{4s}-\Gamma _{c}^{4s}) \\
&\geq &|(K_{a}^{4s}-\Gamma _{a}^{4s})(K_{b}^{4s}-\Gamma
_{b}^{4s}-K_{c}^{4s}+\Gamma _{c}^{4s})|.
\end{eqnarray*}%
The validity of which can be easily proved, noticing that $|K_{\alpha
}^{4s}-\Gamma _{\alpha }^{4s}|\leq 1$.\textbf{\ }

Generally, besides the local one $P_{(s)}^{lc}(ab)$ discussed above, the
measured correlation\textbf{\ }$P_{(s)}(ab)$ includes also the non-local
parts $P_{(s)}^{nlc}(ab)=4\rho _{_{(s)}11}^{nlc}$\textbf{\ (}or\textbf{\ }$%
-4\rho _{_{(s)}22}^{nlc}$) for the Bell cat-states of half-integer spins
because of the nontrivial BP, and then the BI can be violated.

\section{Generalization of spin-parity phenomenon beyond spin-singlet states}

As of yet the investigations of BIs and spin-parity phenomenon have been
restricted on the spin-singlet states only following Bell. It has
demonstrated that the BI holds for all non-product states (i.e. the LMs) and
can be violated by all entangled spin-$1/2$ states \cite{42} 
\begin{equation*}
|\psi _{\frac{1}{2}}\rangle =c_{1}|+,-\rangle +c_{2}|-,+\rangle
\end{equation*}%
where $c_{1}$ and $c_{2}$ are arbitrary normalization constants that $%
|c_{1}|^{2}$ $+|c_{2}|^{2}=1$. We now generate the BI and spin-parity
phenomenon beyond the spin-singlet states. The density operator of LM
corresponding to the arbitrary entangled state $|\psi _{\frac{1}{2}}\rangle $
is 
\begin{equation}
\overset{\wedge }{\rho }_{lc}=|c_{1}|^{2}|+,-\rangle \langle
+,-|+|c_{2}|^{2}|-,+\rangle \langle -,+|  \label{9}
\end{equation}%
while the non-local part is $\overset{\wedge }{\rho }_{nlc}=c_{1}c_{2}^{\ast
}|+,-\rangle \langle -,+|+c_{1}^{\ast }c_{2}|-,+\rangle \langle +,-|$.
Repeating the same calculation the measured correlation probability for the
LM is exactly the same as the singlet case Eq.(4) and thus the BI, CHSH and
Wigner inequalities hold for all LMs given by Eq.(9) in agreement with Ref.%
\cite{42}. For the non-local density-matrix elements we have the same
relation $\rho _{11}^{nlc}=\rho _{44}^{nlc}=-\rho _{22}^{nlc}=-\rho
_{33}^{nlc}$ as Eq.(3), and the non-local correlation is found as 
\begin{equation*}
P_{nlc}(ab)=2\sin \alpha \cos \alpha \sin \theta _{a}\sin \theta _{b}\cos
(\phi _{a}-\phi _{b}+\delta )
\end{equation*}%
where the normalization constants have been parametrized as $c_{1}=\cos
\alpha e^{i\gamma _{1}}$, $c_{2}=\sin \alpha e^{i\gamma _{2}}$ with $\delta
=\gamma _{1}-\gamma _{2}$. The non-local correlation reduces to the singlet
case for $\alpha =-\pi /4+2n\pi $ and $\delta =2n\pi $ with $n$ being
integer. The BI is violated generally by all the entangled states $\overset{%
\wedge }{\rho }=$ $\overset{\wedge }{\rho }_{lc}+$ $\overset{\wedge }{\rho }%
_{nlc}$in agreement also with the conclusion of Ref.\cite{42}$.$

For the spin-$1$ entangled states $|\psi _{1}\rangle =c_{1}|+1,-1\rangle
+c_{2}|-1,+1\rangle $, it is easy to check that the four non-local
density-operator elements are equal as shown in Eq.(5) and the outcome
non-local correlation-probability vanishes $P_{nlc}(ab)=0$. \emph{Thus, both
the Bell and CHSH inequalities will not be violated for all the spin-}$1$%
\emph{\ entangled-states }$|\psi _{1}\rangle $\emph{.}

The generation to spin-$s$ arbitrary Bell cat-states $|\psi _{s}\rangle
=c_{1}|+s,-s\rangle +c_{2}|-s,+s\rangle $ is straightforward in terms of our
formalism. For all the LMs we obtain the same Bell correlation as Eq.(7),
and thus the Bell and CHSH inequalities ought to be satisfied. The
density-matrix element equality Eq.(6) also holds for the general Bell
cat-states and so does the spin-parity phenomenon.

The violation of BIs proved in Ref.\cite{42} for spin-$1/2$ entangled states
was subsequently generated to the arbitrary spins \cite{43}, where two
observable-operators $A$ and $B$ in subspaces are considered as
block-diagonal matrices with each block being an ordinary Pauli matrix in
order to have eigenvalues $\pm 1$. We consider the measuring correlation of
two ordinary spin-$s$ operators $\widehat{s}\cdot \mathbf{a}=s($ $\sin
\theta _{a}$ $\cos \phi _{a}\widehat{s}_{x}$ +$\sin \theta _{a}$ $\sin \phi
_{a}\widehat{s}_{y}+\cos \theta _{a}\widehat{s}_{z})$ and $\widehat{s}\cdot 
\mathbf{b}$ in $2s+1$ dimensional spaces instead. Without the measuring
outcome-restriction one can have only the Bell correlation $\overline{P}%
(ab)=\langle \psi _{s}|\widehat{s}\cdot \mathbf{a}\widehat{s}\cdot \mathbf{b|%
}\psi _{s}\rangle =-s\cos \theta _{a}\cos \theta _{b}$, which can be
verified by the direct calculation of quantum average in the arbitrary Bell
cat-states $|\psi _{s}\rangle $ following Ref.\cite{43}. This is because all
elements of non-diagonal spin-operators vanish i.e. $\langle +s|\widehat{s}%
_{x}|-s\rangle =\langle +s|\widehat{s}_{y}|-s\rangle =0$ for $s\geq 1.$There
is no room for the violation of BIs in the Bell cat-states $|\psi
_{s}\rangle $. However, when measuring outcomes are restricted to the
macroscopic quantum states $|\pm \mathbf{r}\rangle _{s}$ ( $\mathbf{r=a}$, $%
\mathbf{b}$) with eigenvalues $\pm s$, which are more suitable to the
practical experiments, the outcome-independent base vectors of two-particle
measurements Eq.(1) ought to be adopted and the correlation probability is
given by Eq.(2) in the $2\times 2$ dimensional subspaces. Thus the crucial
spin-parity phenomenon is caused surely by the measurement induced geometric
phase, which leads to the violation of BIs only for the half-integer spins.

\section{Conclusions and Discussions}

In terms of the quantum probability average under the assumption of outcome
independence, the Bell correlations are obtained explicitly, with which we
derived analytically the Bell and CHSH inequalities for the Bell models. Our
new observation is that the BIs are not violated for PESs of integer-spins
although they are typically non-local and the violation for the half-integer
spins can be understood as the effect of BP. The relation between the BP and
quantum non-locality in the sense of BI violation has been found. The
spin-parity phenomenon holds for not only spin-singlet but also all
spin-states of bipartite-entanglement as it should be since it resulted only
from the reversal of spin-polarization measurements and is independent of
the initial superposition-coefficients $c_{1}$ and $c_{2}$.

A natural question is whether or not the Wigner inequality can be extended
to the high spin case? The spin-parity phenomenon holds for not only
spin-singlets but also all bipartite-entanglement spin-states as it should
be since it resulted only from the opposite-direction measurement of the
spin polarization and is independent of the initial state Note that here, in
the present Wigner correlations, only one direction of spin-polarization is
involved. Due to such a feature different from those in Bell and CHSH
correlations, the cancellation of the interference effect in relation with
the BP does not exist. The Wigner inequality could be violated even for the
LM of spin-$s$ (with $s\geq 1$) singlets. For example, the two-particle
Wigner correlation for the LM of spin-$1$ singlet is simply $%
P_{(1)}^{lc}(+a,+b)=\rho _{(1)11}^{lc}=\rho _{(1)44}^{lc}$. Consequently, 
\begin{equation*}
P_{(1)}^{lc}(+a,+b)+P_{(1)}^{lc}(+a,+c)=\frac{1}{16}(1+\cos ^{2}\theta
_{a})[2+\cos ^{2}\theta _{b}
\end{equation*}%
\begin{equation*}
+\cos ^{2}\theta _{c}-\ \cos \theta _{a}(\cos \theta _{b}+\cos \theta _{c})],
\end{equation*}%
$\ $and 
\begin{equation*}
P_{(1)}^{lc}(+b,+c)=\frac{1}{16}[1+\cos ^{2}\theta _{b}+\cos ^{2}\theta _{c}
\end{equation*}%
\begin{equation*}
-4\cos \theta _{b}\cos \theta _{c}+\cos ^{2}\theta _{b}\cos ^{2}\theta _{c}].
\end{equation*}%
Now, we can easily check that, for $\theta _{c}=\pi $, $\theta _{a}=\pi /2$,
and $\theta _{b}=0$, the Wigner inequality is violated even for the LM $\hat{%
\rho}_{(1)}^{lc}$.

Our generic arguments of the spin-parity effect could be tested specifically
with the photon-pairs generated by spontaneous parametric down conversion of
light through a BBO crystal. The spin-$1/2$ singlet can be easily
demonstrated with the usual polarization entanglement of photons. And, the
BI for the spin-$1$ singlet can be tested by utilizing the orbital angular
momenta entanglement~\cite{37}, e.g., 
\begin{equation*}
|\psi \rangle _{1}=c_{1}|l_{1}=1,l_{2}=-1\rangle
+c_{2}|l_{1}=-1,l_{2}=1\rangle
\end{equation*}%
of two Laguerre Gaussian mode photons LG$_{\pm 1}$. Such a spin-$1$ singlet
can be produced by applying a Gaussian beam $|\psi _{0}\rangle $ (with $l=0$%
) to the BBO crystal and utilizing the so-called entanglement concentration
technique~~\cite{19}. Next, the photon angular-momentum along arbitrary
directions $\pm \mathbf{a}$ and $\pm \mathbf{b}$ can be engineered and
measured independently by the detectors in each side respectively. Then, the
quantum correlation can be obtained as 
\begin{equation*}
p(a,b)=p_{+a}p_{+b}-p_{+a}p_{-b}-p_{-a}p_{+b}+p_{-a}p_{-b},
\end{equation*}%
where $p_{\pm a}$ denotes the measured probability of angular-momentum
eigenvalues $\pm 1$ along direction $\mathbf{a}$. Finally, the nonviolation
of BI for the spin-$1$ singlet can be thoroughly examined.

\smallskip 
This work was supported in part by the National Natural Science Foundation
of China, under Grants No.11075099, 11275118, 11174373, 90921010, and the
National Fundamental Research Program of China, through Grant No.
2010CB923104.

*jqliang@sxu.edu.cn

\end{document}